\documentclass[prl,twocolumn,showpacs,superscriptaddress,amsmath,amssymb]{revtex4}

\usepackage{graphicx}
\usepackage{dcolumn}
\usepackage{bm}

\def\kbar{\protect\@kbar}
\def\@kbar{%
\relax \bgroup
\def\@tempa{\hbox{\raise.73\ht0
\hbox to0pt{\kern.25\wd0\vrule width.5\wd0
height.1pt depth.1pt\hss}\box0}}%
\mathchoice{\setbox0\hbox{$\displaystyle k$}\@tempa}%
{\setbox0\hbox{$\textstyle k$}\@tempa}%
{\setbox0\hbox{$\scriptstyle k$}\@tempa}%
{\setbox0\hbox{$\scriptscriptstyle k$}\@tempa}%
\egroup}

\begin{document}

\title{Quantized Rotation of Atoms From Photons with Orbital Angular Momentum}

\author{M. F. Andersen}
\author{C. Ryu} 
\author{Pierre Clad\'e}
\author{Vasant Natarajan}
\altaffiliation[Permanent Address: ]{Department of Physics, Indian Institute of Science, Bangalore, India}
\author{A. Vaziri}
\altaffiliation[Present Address: ]{Institut f\"{u}r Quantenoptik und Quanteninformation, Austrian Academy of Sciences, Boltzmanngasse 3, A-1090 Vienna, AUSTRIA}
\author{K. Helmerson}
\author{W. D. Phillips}
\affiliation{Atomic Physics Division,
National Institute of Standards and Technology, Gaithersburg, Maryland 20899-8424, USA}





\date{\today}

\begin{abstract}
We demonstrate the coherent transfer of the orbital angular momentum of a photon to an atom in quantized units of $\hbar$, using a 2-photon stimulated Raman process with Laguerre-Gaussian beams to generate an atomic vortex state in a Bose-Einstein condensate of sodium atoms. We show that the process is coherent by creating superpositions of different vortex states, where the relative phase between the states is determined by the relative phases of the optical fields. Furthermore, we create vortices of charge 2 by transferring to each atom the orbital angular momentum of two photons.
\end{abstract}

\pacs{03.75.Lm, 03.75.-b, 32.80.-t, 42.50.Vk}   
\maketitle

Light can carry two kinds of angular momentum: Internal or spin angular momentum (SAM) associated with its polarization and external or orbital angular momentum (OAM) associated with its spatial mode \cite{LLbook}. A light beam with a phase singularity, e.g., a Laguerre-Gaussian (LG) beam, has  a well-defined OAM along its propagation axis \cite{allen92}.  Beams with phase singularities have only recently been generated \cite{tamm90, soskin90, norm92}, and are now routinely created so as to carry specific values of OAM \cite{OAMbook, allen93}.

Interaction of light with matter inevitably involves the exchange of momentum. For linear momentum (LM), the mechanical effects of light range from comet tails to laser cooling of atoms. The transfer of optical SAM to atoms has been studied for over a century \cite{zeeman1897}, and the mechanical effect of SAM on macroscopic matter was first demonstrated 70 years ago in an experiment where circularly polarized light rotated a birefringent plate \cite{beth36}. More recently, the mechanical effects of optical OAM on microscopic particles and atoms have been investigated \cite{OAMbook}. SAM and OAM of light has been used to rotate micron-sized particles held in optical tweezers \cite{halina95, halina96, padgett97}. The forces on atoms due to optical OAM \cite{footnote1} have also been investigated theoretically \cite{OAMbook} and experimentally. In one series of experiments \cite{tabosa99}, a diffraction grating was created in an atomic cloud, such that diffraction of a Gaussian (G) beam generated a light beam carrying OAM. Another experiment \cite{kozuma03} used a technique similar to phase imprinting \cite{machida00} to generate a light beam with OAM.  In both cases, mechanical OAM was likely transferred to the atomic clouds, but not directly observed.  (Such an observation would have been difficult, since the atomic clouds were incoherent, thermal samples.) No experiment has demonstrated the quantized transfer of the OAM of a photon to an atom.

An atomic gas Bose-Einstein condensate (BEC) allows the study of macroscopic quantum states. For example, BEC superfluid properties can be explored using vortex states (macroscopic rotational atomic states with angular momentum per atom quantized in units of $\hbar$). The many-body wavefunction of the BEC is very well approximated by the product of identical single-particle wavefunctions, so for a BEC in a vortex state, each particle carries quantized OAM. The first generation of a vortex in a BEC used a "phase engineering" scheme involving a rapidly rotating G laser beam coupling the external motion to internal state Rabi oscillations \cite{cornell99, williams99}. Later schemes included mechanically stirring the BEC with a focused laser beam \cite{dalibard00} and "phase imprinting" by adiabatic passage \cite{machida00, ketterle04}. However, transfer of OAM from the rotating light beams in these earlier schemes is not well-defined.

Here, we report the direct observation of the quantized transfer of well-defined OAM of photons to atoms. Using a 2-photon stimulated Raman process, similar to Bragg diffraction \cite{kozuma99}, but with a LG beam carrying OAM of $\hbar$ per photon, we generate an atomic vortex state in a BEC. Over the past decade, numerous papers \cite{schemes, dowling05} proposed generating vortices in a BEC using stimulated Raman processes with configurations of optical fields that provide OAM, such as LG beams. Our experiment is the first realization of this technique, but differs from the proposals in that we do not change the internal atomic states; instead we change the LM state transferring OAM in the process.
Furthermore, we demonstrate that the process is coherent by creating superpositions of different vortex states where the relative phase between the states is determined by the relative phases of the optical fields. Our process represents both a new and well-controlled way of creating a vortex state in a BEC and a new tool for the coherent control of the OAM of atomic samples, complementing existing tools for LM and SAM.

The set of Laguerre-Gaussian modes ($\rm{LG}^{{\it l}}_{{\it p}}$) defines a possible basis set to describe paraxial laser beams \cite{siegman, barnett02}. 
The indice $\it{l}$ is the winding number or charge (the number of times the phase completes $2\pi$ on a closed loop around the propagation axis) and $\it{p}$ is the number of radial nodes for radius $\rho>0$. Each photon in the $\rm{LG}^{{\it l}}_{{\it p}}$ mode carries $l\hbar$ of OAM along its direction of propagation \cite{allen92}. In contrast, SAM can only carry $\hbar$ of angular momentum per photon. We use a $\rm{LG}^{1}_{0}$ mode, where the electric field amplitude in polar coordinates at the beam waist varies as,
\begin{equation}
{\rm{LG}}^{1}_{0}\left(\rho , \phi \right)=\frac{2}{\sqrt{\pi}}
\frac{1}{w_{0}^{2}} \rho \exp \left( - \frac{\rho ^2}{w_{0}^{2}}
\right)\exp \left( i \phi \right), \label{lg}
\end{equation}
and the peak-to-peak diameter is $\sqrt{2}w_{0}$. The light is linearly polarized and carries no net SAM. 

Figure 1 shows a stimulated Raman scheme using counter-propagating $\rm{LG}^{1}_{0}$ and G beams. An atom of mass $M$, initially at rest, absorbs a $\rm{LG}^{1}_{0}$ photon and stimulatedly emits a G photon, acquiring $2\hbar k$ of LM ($k = 2\pi/\lambda$ with $\lambda$ the photon wavelength). As with resonant Bragg diffraction with two G beams, the frequency difference between the two beams, $\delta \nu$, is $4E_{r}/h = 4\nu_{r}$, where $E_{r} = {(\hbar k)^2}/2M$ is the recoil energy \cite{kozuma99}. In addition to LM the atoms pick up the OAM difference between the two photons. The additional energy due to the rotation is small and, for the pulse durations used in this experiment, does not affect the resonance condition \cite{footnote2}.
\begin{figure}
\includegraphics*[scale=0.091, bb= 0 00 2700 1300]{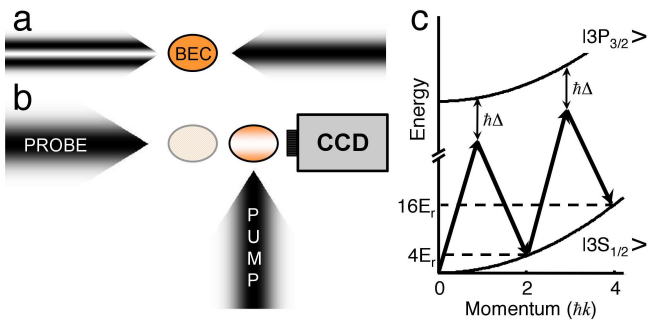}
\caption{Schematic of the experiment. (a) Counter-propagating $\rm{LG}^{1}_{0}$ and Gaussian laser beams, with the same linear polarization and a variable frequency difference of $\delta \nu$, are applied to a BEC. (b) The atoms that have undergone the Raman transition (right cloud) have separated from those that did not (left cloud). A spatially localized ÒpumpÓ beam enables independent imaging of each cloud by absorption of a probe beam propagating along the direction of LM transfer. (c) Diagram illustrating energy and LM conservation of the 2-photon Raman process for one and two consecutive pulses.}
\label{schematic}
\end{figure}
The LM transferred by Bragg diffraction can be viewed as the result of the diffraction of atoms from a moving sinusoidal optical dipole potential generated by the interference of the counter-propagating G beams. Here, the optical dipole potential generated by interference of the counter propagating $\rm{LG}^{1}_{0}$ and G beams is not sinusoidal, but, due to the radial intensity profile and the helical phase of the $\rm{LG}^{1}_{0}$ beam, the dipole potential generated is corkscrew-like. Diffraction off this corkscrew potential produces a rotating state. The potential is the atom-optics analogue of a phase hologram, and one could generate any desired two-dimensional atomic state using a suitable hologram. 

The experiment begins with a BEC of 1-2$\times10^{6}$ sodium atoms in the $|3S_{1/2}, F=1, m_F =-1\rangle$ state prepared as described in \cite{kozuma99}.  The atoms are  confined in a triaxial time-orbiting potential (TOP) magnetic trap \cite{kozuma99} with trapping frequencies of $\nu_z = \sqrt{2}\nu_y = 2\nu_x = 40$ Hz (gravity along $z$) yielding a condensate with Thomas-Fermi radii of $21, 30$ and $42$ $\mu$m, respectively. A G laser beam, detuned from the D2 line ($\lambda = 589.0$ nm) by $\Delta = -1.5$ GHz ($\approx$$150$ linewidths, enough to prevent any significant spontaneous photon scattering), is split into two beams that pass through separate acousto-optic modulators (AOMs) in order to control their frequency difference $\delta \nu$. One of the beams diffracts from a blazed transmission hologram \cite{soskin90, norm92, zeilinger01} generating a $\rm{LG}^{1}_{0}$ mode. The $\rm{LG}^{1}_{0}$ beam, with a power of $1.5$ $\mu$W and $w_{0} = 85$ $\mu$m at the BEC, propagates along $x$. The G beam, with power $18$ $\mu$W and $1/e^{2}$ radius $w_{0} \approx175$ $\mu$m, propagates along $-x$. We apply these beams to the trapped atoms as a square pulse and then turn off the trap. After 6 ms time-of-flight (TOF), during which the atoms propagate ballistically, we image the released atoms by absorption of a probe beam resonant with the $|3S_{1/2}, F=2\rangle$ to $|3P_{3/2}, F=3\rangle$ transition. During imaging the atoms must be optically pumped from the initial $|3S_{1/2}, F=1\rangle$ state into the $|3S_{1/2}, F=2\rangle$ state by a pump beam resonant with the  $|3S_{1/2}, F=1\rangle$ to $|3P_{3/2}, F=2\rangle$ transition. Atoms with LM $2\hbar k$ from the Raman process will separate spatially during the TOF from atoms still at rest (see Fig. 1b). We use a focused pump beam spatially localized along $x$ to selectively image clouds of atoms in different LM states using a probe beam propagating along $x$, the axis of propagation of the $\rm{LG}^{1}_{0}$ beam. 

Figure 2a shows an image of a cloud that has undergone the Raman process with $\delta \nu \approx4\nu_{r} \approx 100$kHz, where the vortex core is observed as a hole in the middle of the cloud. (A hole in the atomic density distribution without rotation would fill in during the TOF expansion.) For a $130$ $\mu$s pulse a maximum transfer efficiency of $53\%$ was achieved. The transfer is limited by the spatial mismatch between the (toroidal shape) rotating state and the (inverted parabolic shape) initial BEC; transfer of the entire BEC, in this situation, is impossible. The size of the Raman laser beams, somewhat larger than the BEC size, was chosen to give good transfer efficiency, but was not carefully optimized.  The power is chosen to give a $\pi$-pulse for a duration in the Bragg regime \cite{kozuma99}  that is shorter than the time scale of the trap oscillation, meanfield energy, and Doppler broadening.
\begin{figure}
\includegraphics*[scale=0.42, bb= 10 110 600 700]{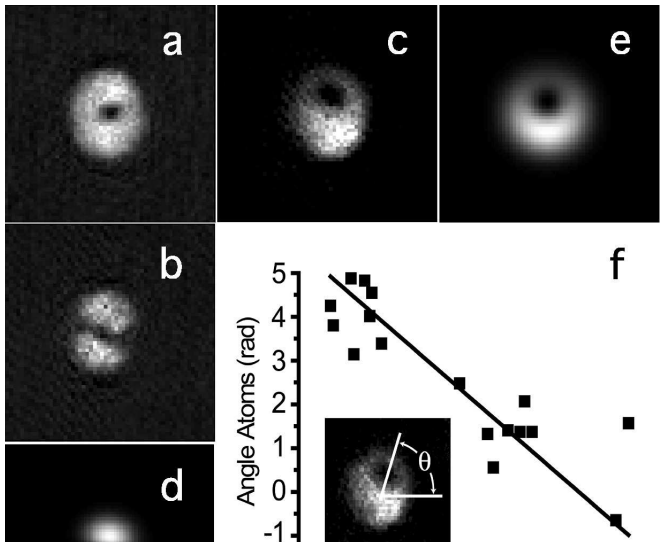}
\caption{(a) Absorption image of a cloud that has undergone the Raman transition, taken along the axis of the $\rm{LG}^{1}_{0}$ beam. The vortex core is seen as a hole in the cloud. (b) Interference between left and right rotating clouds. (c) Interference pattern between a non-rotating and a rotating cloud, showing a displaced hole. (d) Calculated interference pattern between left and right rotating states. (e) Calculated interference pattern between a non-rotating and a rotating state. (f) Angle of the hole in the interference pattern between rotating and non-rotating atomic states as a function of the rotation angle of the optical interference pattern between the $\rm{LG}^{1}_{0}$ and co-propagating Gaussian beams. The straight line (to guide the eye) has slope $-1$. Inset: Image of the atomic interference between a rotating and non-rotating cloud. The hole is displaced from the center and its angular position $\theta$ depends on the relative phase between the interfering states.}
\label{fiveplusgraph}
\end{figure}
To measure the angular momentum transferred to the atoms, we perform an interferometric measurement using three optical pulses. The first pulse, consisting of the $\rm{LG}^{1}_{0}$ beam and the counter-propagating G beam ($\rm{LG}^{1}_{0}/G$ pulse), is $30 $ $\mu$s long and with $\delta \nu \approx4\nu_{r}$ transfers about $20\%$ of the atoms to a state with LM $2\hbar k$ and OAM $+\hbar$. The same two beams are used in the second pulse, $60 $ $\mu$s long, but with $\delta \nu \approx-4\nu_{r}$, which transfers about $40 \%$ of the remaining atoms to a state with LM $-2\hbar k$ and OAM $-\hbar$. The third pulse (G/G pulse) is resonant for a second order (4 photon) Raman process between states with momenta $-2\hbar k$ and $+2\hbar k$ \cite{kozuma99}. This pulse is $100 $ $\mu$s long (chosen empirically to produce high contrast interference) and is produced by replacing the $\rm{LG}^{1}_{0}$ beam with a second G beam with $\delta \nu = 0$, $w_{0} \approx200 $ $\mu$m and power of $8 $ $\mu$W. There is essentially no delay between the pulses so that atoms with different momenta remain well overlapped spatially during the pulse sequence (clouds with LM difference $2\hbar k$ separate in 1 ms). Fig. 2b is an image of one of the interfering clouds after the three pulses, and corresponds to the superposition of two clouds with OAM $\pm \hbar$ (Fig. 2d), which has average OAM of zero. Since each diffracted atom has absorbed or been stimulated to emit one $\rm{LG}^{1}_{0}$ photon, the interference pattern confirms that each $\rm{LG}^{1}_{0}$ photon transfers $\hbar$ OAM to each atom. Although interference has previously been used to observe vortex states \cite{cornell99, dalibard01, ketterle01}, this is the first interference between independently generated, overlapping counter-rotating vortex states \cite{footnote3}.

A stimulated Raman process is coherent. The phase difference of the laser beams determines the phase of the diffracted, rotating cloud. To confirm this we perform a two-pulse experiment. The first pulse is a $30$ $\mu$s $\rm{LG}^{1}_{0}$/G pulse with $\delta \nu \approx 4\nu_{r}$, which diffracts atoms into the $2\hbar k$ LM state with $\hbar$ OAM. The second pulse is a G/G pulse with $\delta \nu \approx 4\nu_{r}$, which also couples the same two LM states ($0$ and $2\hbar k$) but without changing the OAM. Fig. 2c is an image of the $2\hbar k$ cloud from the two-pulse sequence. The off-centered hole results from the interference between a state rotating with OAM $\hbar$ and a non-rotating state. The direction in which the hole is displaced is determined by the phase between the two states 
\cite{cornell99}, which is determined by the relative phase differences of the two Raman pulses. We directly measure this relative phase difference by imaging the interference pattern of the $\rm{LG}^{1}_{0}$ and the co-propagating G beams, since both Raman pulses use the same counter-propagating G beam \cite{footnote4}. This measures the relative position of the corkscrew and sinusoidal diffractive structures generating the two interfering clouds. In Fig. 2f the measured phase of the atomic interference is plotted as a function of the measured relative phase difference of the Raman beams, for 18 consecutive realizations of the experiment. They are correlated, as expected. Hence atoms can be put into any desired coherent superposition of rotational states by controlling the relative phases of the Raman beams.

We generate vortices of higher charge by transferring to each atom the angular momentum from several $\rm{LG}^{1}_{0}$ photons (see Fig. 1c). A $30$ $\mu$s $\rm{LG}^{1}_{0}/G$ pulse with $\delta \nu \approx 4\nu_{r}$ first transfers $18 \%$ of the atoms into the singly charged vortex state with LM $2\hbar k$. A second $\rm{LG}^{1}_{0}/G$ pulse, $70$ $\mu$s long with $\delta \nu \approx 12\nu_{r}$, transfers $80 \%$ of the atoms in the $2\hbar k$ state into a doubly charged vortex state with LM $4\hbar k$. Fig. 3a is an image of this state. (A doubly charged vortex has previously been created in a BEC using "phase engineering" \cite{haljan02} and Òphase imprintingÓ  \cite{ketterle04}, respectively.) To verify that this is a doubly charged vortex, we apply a third G/G pulse, $40$ $\mu$s long with $\delta \nu \approx 8\nu_{r}$, which couples states with momentum $0$ and $4\hbar k$ via a second order Raman process \cite{kozuma99}. Fig. 3b is an image of the $4\hbar k$ cloud generated by the three pulses, taken after 6 ms TOF. It corresponds to the interference between a non-rotating cloud and a cloud with OAM $2\hbar$ (see Fig. 3c), as expected. 
\begin{figure}
\includegraphics*[scale=2.67, bb=  0 0 100 30]{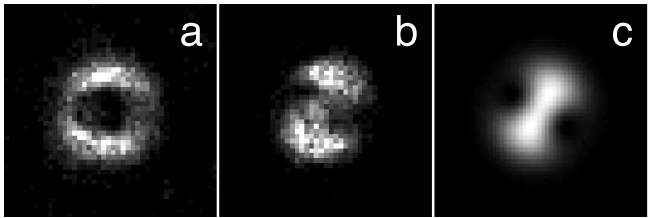}
\caption{(a) Absorption image of the doubly charged ($+2\hbar$) vortex cloud. The core is seen to be larger than for the single charged state of Fig. 2a. (b) Absorption image of the cloud resulting from the interference between a doubly charged state and a non-rotating state. (c) Calculated interference pattern between non-rotating and doubly charged rotating state.}
\label{doublevortex}
\end{figure}
Our experiments directly demonstrate that the OAM of a photon is transferred coherently to an atom in quantized units of $\hbar$. Although we transferred LM in addition to OAM, in order to achieve good discrimination between initial and final states of the Raman process because of the small rotational energy, in some situations it might be desirable to generate rotational states with no net LM. For example, atoms in a ring trap \cite{stamper-kurn05, stringari06} could be induced to rotate, resulting in a persistent current of atoms. This could be accomplished by using an initial Bragg diffraction pulse to put atoms in a non-zero LM state from which they could subsequently be transferred to a rotational state with zero LM. Alternatively one could use co-propagating beams and drive transitions between different internal states in the atoms as proposed in \cite{schemes, dowling05}. If longer pulse lengths were used, it may be possible to directly induce a rotation of the condensate without changing the internal state or transferring linear momentum; however, such a process may be strongly suppressed \cite{footnote5} since, in the Thomas-Fermi regime, the rotational energy is much less than the mean-field interaction energy.

In summary, we've demonstrated a new tool to generate arbitrary superpositions of atomic rotational states, which with tools for controlling LM and internal states enables total control of an atom. 
Applications range from generating superflow and superposition of macroscopic (Schr\"{o}dinger cat) states in atomic vapors to quantum information \cite{dowling05}, for example in quantum repeaters where the flying qubits are photons with OAM \cite{zeilinger01}.

\begin{acknowledgments}
This work was partially supported by ONR, NASA and ARDA. We thank A. Aspect and S. Barnett for helpful discussions. A.V. acknowledges partial support by the European Commission under the Integrated Project Qubit Applications (QAP) funded by the IST  directorate as Contract Number 015848.
\end{acknowledgments}


\begin{references}
\bibitem{footnote1} In contrast to the situation for a macroscopic object, transfer of SAM will not result in a mechanical rotation of an atom about the axis of propagation of the photon, but transfer of the OAM will produce a rotation.
\bibitem{footnote2} In our experiment, the rotational energy is less than $1$ Hz, which is much smaller than the Fourier width of the Bragg pulse. Therefore, choosing the frequency difference between the two beams to satisfy the Bragg resonance condition of $4\nu_{r}$ ($100$ kHz for sodium)\cite{kozuma99} is sufficient to resonantly transfer both LM and OAM to the atoms.
\bibitem{footnote4} We measure the relative phase difference 10 ms after the diffracting pulses. The phase difference of the Raman beams is primarily determined by changes in the optical path, which mostly depends on mirror vibrations. We've confirmed that on this timescale, such phase changes are negligibly small.
\end{references}
\end{document}